\documentstyle{article}

\title{ Possible evidence for the mass shift of $\eta$' meson 
        at finite temperature }

\author{S. Hioki \\ Department of Physics, Hiroshima University, \\
                    Higashi-Hiroshima 739, Japan\\
		TEL: +81-824-24-7476\\
		FAX: +81-824-24-0717\\
		e-mail: hioki@sci.hiroshima-u.ac.jp 
	}

\begin{document}
\large

\date{}
\maketitle
\baselineskip 7mm
\vspace*{1cm}

\begin{center}          Abstract              \end{center}

 Topological charge distributions of SU(3) gauge theory 
at finite temperature are calculated on the lattice with high statistics
in a manner of free from the uncertainty of $\beta$-function.
 Clear temperature dependence of the topological charge distribution 
is obtained.
 Combining the present result with the recent study about the relation
between the masses of $\eta$ meson system and the gauge field topology,
we obtain 
$m_{\eta'}(T)/m_{\eta'}(0) = 
0.86 \pm 0.02$ and $0.69 \pm 0.01$ 
for $T=0.75 T_c$ and $T=0.93 T_c$, respectively.
The result shows the first clear evidence of the mass shift of
$\eta'$ meson at finite temperature.

\clearpage

\section{Introduction}

 It is a common knowledge that at sufficiently high temperature,
the hadronic system may undergo a phase transition to the QGP 
(quark-gluon-plasma) phase.
 What is a signature or a precursor of the QGP is a long standing question.
 Although a lot of possibilities have been posted so far,
there is no definite answer yet \cite{QM}.
 Among them the mass shift of hadrons might be an promising candidate
for the precursor of QGP if this phenomena is in the case.
\cite{Pisa82,Pisa84,FTmass,Hashi}

 This is because the mass shift, itself, can indicate that the system
is very close to the critical region without a precise quantitative
argument. 

 In this paper, we focus on the neutral pseudo-scalar meson system 
whose mass spectrum at zero temperature can be well explained
by the linear sigma model incorporated with the axial U(1) anomaly 
\cite{Pisa84}.
 It is also well known that the axial U(1) anomaly is caused by 
the topological fluctuations of the system,
\begin{equation}
 \partial_{\mu} j_{\mu}(x) \propto  Q(x),
\end{equation}
where $j_{\mu}$ is the flavor singlet axial vector current and
$Q(x)$ is the topological charge density.

 Recently the relation between the singlet-octet mass difference of 
the neutral pseudo-scalar meson system and the topology 
has been analyzed in detail and the strong correlations of these two
has been suggested \cite{Fuku95}. 

 From these results mentioned above, if there exist the temperature
dependence in the topological sector of the theory, it might be
reflected by the possible change of the mass spectrum.

 The organization of this paper is as follows.
 In the next section 2, 
we briefly mention our model of the neutral pseudo-scalar meson system
as well as it's relation with the gauge field topology.
 The lattice results of the topological charge distributions 
at finite temperature is given in section 3.
 In the section 4, we will discuss the possible shift of the masses.
 The section 5 is devoted to the conclusion.

\section{Model of the neutral pseudo-scalar meson system}

 For the model of the neutral pseudo-scalar meson system, we adopt the 
following mass matrix incorporated with the axial U(1) anomaly
which was introduced by Pisarski and Wilczek \cite{Pisa84}
and was able to reproduce the real mass spectrum at zero temperature.

\begin{eqnarray}
M^2_{\pi^0 \pi^0}&=&(m_u + m_d) {v \over f^2_{\pi}}, \nonumber\\
M^2_{\pi^0 \eta} &=&{(m_u - m_d) \over \sqrt{3}}  
                    {v \over f^2_{\pi}}, \nonumber\\
M^2_{\eta \eta}  &=&{(m_u + m_d +4 m_s) \over 3} 
                    {v \over f^2_{\pi}}, \\
M^2_{\pi_0 \eta'}&=&\sqrt{2\over 3} (m_u - m_d )
                    {v \over f_{\pi} f_{\eta'}}, \nonumber\\
M^2_{\eta \eta'} &=&{\sqrt{2} \over 3} (m_u + m_d -2 m_s)
                    {v \over f_{\pi} f_{\eta'}}, \nonumber\\
M^2_{\eta' \eta'}&=&{2 \over 3} (m_u + m_d + m_s)
                    {v \over f^2_{\eta'}} + K, \nonumber
\end{eqnarray}
where $M^2$ is the (mass)$^2$ matrix for the $\pi_0 - \eta - \eta'$ system,
$m_i$ is the quark mass for i-quark (i=up, down and strange),
$f_j$ is the decay constant for j-meson (j=$\pi_0, \eta$ and $\eta'$),
$v$ is the strength of the chiral condensate 
($-v=<u\bar u>=<d\bar d>=<s\bar s>$) and $K$ is the free parameter
representing the effects of the axial U(1) anomaly.

 In order to solve this equation at finite temperature, the
temperature dependence of all parameters in the above equation
should be given.
Pisarski and Wilczek have considered only the
$K$ effect of the system.\cite{Pisa84} 
 Although there have been several model works which suggest the possible
change of other parameters like $f_{\pi}$ and $v$ at finite temperature,
the recent lattice calculation indicates that the temperature dependence
of these quantities seems very small up to very close to the critical 
temperature \cite{Learmann94}.

 So in this paper we also restrict ourselves on the $K$ effect of
the system.

 Let us first consider the case that the flavor SU(3) symmetry
is unbroken, i.e., $m_u = m_d = m_s$.
 In this case, $K$ is nothing but the mass difference $m_0$ between
singlet and octet meson which is directly related the topological charge
by the result of ref.\cite{Fuku95}.
Then once we know the
topological charge distribution $\rho(Q)$, we can estimate $K$ as,
\begin{equation}
\label{eq:KT}
K = {\bar m_0} \equiv \int dQ \tilde\rho (Q) m_0(Q),
\end{equation}
where
\begin{equation}
\label{eq:normrho}
\tilde\rho (Q) = \rho (Q) / \int dQ \rho (Q)
\end{equation}
is the normalized topological charge distribution.

 In eq.2, it is also clear that $K$ represents the strength of the axial
U(1) anomaly which is connected with the gauge field topology via eq.1.
 So it is rather plausible to assume that the $K$ is independent of the
flavor SU(3) symmetry.
 Under this assumption,
above equation becomes valid  even in the flavor symmetry broken phase.

 Next we assume the singlet-octet mass difference $m_0$ is purely
due to the gauge field topology, i.e.,
all temperature effect to $m_0$ should be included through the change
of the topological charge distribution $\rho (Q)$.
 Then we can use eq.3 even at finite temperature.

\section{Topological charge distribution at $T\neq 0$}

 When we evaluate eq.3 at finite temperature on the lattice,
there are two restrictions at this stage;
\begin{itemize}
\item $m_0(Q)$ was obtained at only one $\beta(=6/g^2)$ \cite{Fuku95},
\item unknown non-perturbative behavior of $\beta$-function.
\end{itemize}

 In order to get rid off these restrictions, we have decided to make
simulation at fixed $\beta(=6/g^2)$.
 In this case, $T\neq 0$ can be realized by changing $N_t$ 
(= temporal extent of the lattice) through $T=1/N_t a$.
( $a$ is the lattice spacing for all directions. )

 The simulation is done on $16^3 \times 16$,\ $8$ and $6$ lattice in SU(3)
gauge at $\beta=5.89$.
 These lattices correspond to the temperature $T\simeq 0,\ 0.75 T_c$
and $0.93 T_c$, respectively, by the fact that the critical $\beta$ on
$16^3\times 6$ lattice is 
$\beta_c\sim 5.894$ \cite{Nt6} 
and by using $a^{-1}$ vs. $\beta$ data by QCDTARO
Collaboration \cite{QCDTARO93}.
 For the SU(3) simulation code in this analysis, we use QCDMPI \cite{QCDMPI}.

 After the thermalization of 1000 pseudo-heat-bath sweeps,
the measurement is done every 100 sweeps. Number of configurations used are
3500, 1500 and 3200 for $T\simeq 0$, $T=0.75 T_c$ and  $T=0.93 T_c$,
respectively.
 To extract the topological charge $Q$, we adopt the cooling method
\cite{cooling}.

 The result of $\tilde\rho (Q)$ is shown in Fig.1.
 In this figure, clear temperature dependence of $\tilde\rho (Q)$ is seen.
 The distribution becomes narrower as the temperature gets close to
the critical temperature.
 This feature can be understood as the partial restoration of the axial
U(1) symmetry at finite temperature.

 The change of the topological charge distribution $\rho (Q)$ 
at finite temperature should be treated carefully. 
If we assume the distribution is Gaussian,
$\rho (Q) \propto \exp(-\alpha Q^2)$, the constancy of the topological
susceptibility in the confinement phase \cite{Q2} requires,
\begin{equation}
\label{eq:alphaT}
 \alpha  \propto N_t^{-1}
\end{equation}
at fixed $\beta$.

 In this sense, the distribution becomes narrower as the temperature
increases ($N_t$ decreases). 
 So this factor (eq.\ref{eq:alphaT}) should be taken into account for the
quantitative study in the next section.

\section{Temperature dependence of $K$ and $m_{\eta}$}

 For the parameterization of $m_0(Q)$, we adopt two different fits of the result 
in ref.\cite{Fuku95}.
 One is the least square linear fit (fit A);
\begin{equation}
  m_0(|Q|) = c_1 |Q| + c_2,
\end{equation}
with $c_1=0.08$ and $c_2=0.20$ in MeV units \cite{Fuku95}.
 It is surprising that this linear fit seems very well as you can see
in ref.\cite{Fuku95}.
In this fit $c_2$ is finite. However it is natural for $c_2$ to be zero
at $|Q|$=0 as far as the flavor SU(3) symmetry is unbroken.
So we make another fit (fit B);
\begin{equation}
  m_0(|Q|) = c_1' |Q|\ \ \ \ \ \ \,
\end{equation}
with $c_1'=0.16$,
such that the average mass splitting is almost equal for these two fits.

 Since $\beta$ in this simulation is different from $\beta$ in \cite{Fuku95},
we can not fix the absolute scale without the knowledge of 
the $\beta$-function of SU(3) gauge theory.
 So we concentrate on the mass ratio normalized to the zero temperature value
which is independent of this $\beta$ discrepancy as far as the fits
A and B are concerned.

 When we calculate $K(T)$ in eq.\ref{eq:KT}, $N_t$ effect in eq.\ref{eq:alphaT}
should be eliminated.
 To do this, in this paper, we normalize $K(T)$ by its Gaussian value,
\begin{equation}
\label{eq:KTnew}
  K(T) = { K(T)_{\rm lattice} \over K(T)_{\rm gaussian} }.
\end{equation}

 Combining $m_0(Q)$ and eq.3, we then obtain the temperature dependence
of $K$ parameter. 
 The result is shown in Fig.2.
 We can see the dramatic decrease of this quantity as the temperature
gets close to $T_c$.

 Finally we solve the eigenvalue problem of eq.2 using this $K(T)$.
 Other parameters are taken from ref.\cite{Pisa84}.
 The result for $m_{\eta}(T)$ and $m_{\eta'}(T)$ are shown in Fig.3.
 In these figures, only statistical error coming from
the topological charge measurement has been taken into account.
 Although systematic error coming from above 2 fits (fit A and fit B)
 is not so small,
 mass decrease near $T_c$ is seen for both $\eta$ and $\eta'$.
 ($m_{\pi}$ does not show any noticeable change in this calculation
  as in ref.\cite{Pisa84}.)
 Especially for $\eta'$, clear mass shift can be seen
(in the case of fit B),
\begin{eqnarray}
{m_{\eta'}(T) \over m_{\eta'}(T=0)} &=& 0.86 \pm 0.02 (T=0.75T_c)\\
                                     &=& 0.69 \pm 0.01 (T=0.93T_c)
\end{eqnarray}
 This feature can also be understood as the direct reflection of the
weakenness of the axial U(1) anomaly. 
 From the present result,
 if we measure this temperature dependence in the experiment,
it could be a possible precursor and/or thermometer of the 
QGP phase transition.

\section{Conclusion}

 We have analyzed the temperature dependence of the topological charge
distributions of the SU(3) gauge theory.
 Measurement has been performed on the lattice at fixed $\beta(=6/g^2)$ which
enable us to evaluate the temperature dependence of the physical 
quantity free from the uncertainty of $\beta$-function.
 Although the systematic error coming from the uncertainty of $m_0(Q)$
is not so small, we find the clear mass shift of $\eta'$ meson
near the critical temperature.
 This is a first clear evidence of the mass shift of $\eta'$ meson
at finite temperature.

 There are several assumptions in this paper.
 Although these assumptions seem plausible at this stage, 
the detail investigation are highly expected in order to confirm 
the present result.
 Calculations with dynamical
quark effect should be performed in future because the dynamical quarks
may play a crucial role on this problem.

\section*{Acknowledgements}
 
 Numerical calculations for the present work have been carried out 
on Intel Paragons 
at INSAM (Institute for Numerical Simulations and Applied Mathematics)
in Hiroshima University, at Information Processing Center of Okayama
University of Science and at National Aerospace Laboratory.
 This work is supported by the Grant-in-Aid for Scientific Research 
of the Ministry of Education No.07740223 and No.08640379.

\clearpage

\section*{figure caption}
\begin{itemize}
\item Figure 1\\
	Topological charge distributions $\tilde\rho(Q)$ 
	at (a) $T\simeq 0$, (b) $T=0.75 T_c$ and (c) $T=0.93 T_c$.

\item Figure 2\\
	$K(T)$ normalized to the zero temperature value.
	$\bullet$ is for $m_0(|Q|) = c_1 |Q| + c_2$ fitting and
	$\circ$ is for $m_0(|Q|) = c_1' |Q|$ fitting.
	Errorbars represent only statistical errors.

\item Figure 3\\
	(a) $\eta$' and (b) $\eta$ mass vs. temperature normalized to 
	the zero temperature value
	using the $m_0(|Q|)$ result by 
	Fukugita et al.\protect\cite{Fuku95},
	where $m_0$ is the $\eta$'-octet mass splitting.
	$\bullet$ is for $m_0(|Q|) = c_1 |Q| + c_2$ fitting and
	$\circ$ is for $m_0(|Q|) = c_1' |Q|$ fitting.
	Errorbars represent only statistical errors.

\end{itemize}

\clearpage

\begin{figure}[t] 
\unitlength 2.0mm
\begin{picture}(60,60)(-10,-10)
\def\xw{60.000000} \def\yw{40.000000}
\put(22,-20){\Huge Figure 2}
\put(26,-8){\huge $T/T_c$}
\put(-10,43){\huge $K(T)/K(0)$}
\def\errorbar2#1#2#3#4#5#6{
\put(#1,#2){\line(0,1){#3}}
\put(#1,#2){\line(0,-1){#3}}
\put(#4,#5){\line(1,0){1}}
\put(#4,#6){\line(1,0){1}} }
\errorbar2{0.000000}{40.000000}{0.000000}{-0.500000}{40.000000}{40.000000}
\errorbar2{45.000000}{40.000000}{0.360000}{44.500000}{40.360000}{39.640000}
\errorbar2{55.799999}{35.239998}{0.320000}{55.299999}{35.559998}{34.919998}
\put(0.000000,40.000000){\circle*{1.000000}}
\put(45.000000,40.000000){\circle*{1.000000}}
\put(55.799999,35.239998){\circle*{1.000000}}
\def\errorbar2#1#2#3#4#5#6{
\put(#1,#2){\line(0,1){#3}}
\put(#1,#2){\line(0,-1){#3}}
\put(#4,#5){\line(1,0){1}}
\put(#4,#6){\line(1,0){1}} }
\errorbar2{0.000000}{40.000000}{0.000000}{-0.500000}{40.000000}{40.000000}
\errorbar2{45.000000}{27.320002}{0.280000}{44.500000}{27.600002}{27.040002}
\errorbar2{55.799999}{14.440001}{0.120000}{55.299999}{14.560001}{14.320001}
\put(0.000000,40.000000){\circle{1.000000}}
\put(45.000000,27.320002){\circle{1.000000}}
\put(55.799999,14.440001){\circle{1.000000}}
\put(0.000000,0){\line(0,1){1}}
\put(12.000000,0){\line(0,1){1}}
\put(24.000000,0){\line(0,1){1}}
\put(36.000000,0){\line(0,1){1}}
\put(48.000000,0){\line(0,1){1}}
\put(60.000000,0){\line(0,1){1}}
\put(0.000000,0){\line(0,1){1}}
\put(-2.200000,-3.5){\huge 0.0}
\put(9.800000,-3.5){\huge 0.2}
\put(21.800000,-3.5){\huge 0.4}
\put(33.800000,-3.5){\huge 0.6}
\put(45.800000,-3.5){\huge 0.8}
\put(57.800000,-3.5){\huge 1.0}
\put(1.100000,-3.5){\huge }
\put(0,0.000000){\line(1,0){1}}
\put(\xw,0.000000){\line(-1,0){1}}
\put(0,8.000000){\line(1,0){1}}
\put(\xw,8.000000){\line(-1,0){1}}
\put(0,16.000000){\line(1,0){1}}
\put(\xw,16.000000){\line(-1,0){1}}
\put(0,24.000000){\line(1,0){1}}
\put(\xw,24.000000){\line(-1,0){1}}
\put(0,32.000000){\line(1,0){1}}
\put(\xw,32.000000){\line(-1,0){1}}
\put(0,40.000000){\line(1,0){1}}
\put(\xw,40.000000){\line(-1,0){1}}
\put(0,0.000000){\line(1,0){1}}
\put(\xw,0.000000){\line(-1,0){1}}
\put(-5.700000,-1.000000){\huge 0.0}
\put(-5.700000, 6.900000){\huge 0.2}
\put(-5.700000,14.900000){\huge 0.4}
\put(-5.700000,22.900000){\huge 0.6}
\put(-5.700000,30.900000){\huge 0.8}
\put(-5.700000,38.900000){\huge 1.0}
{\linethickness{0.25mm}
\put(  0,  0){\line(1,0){\xw}}
\put(  0,\yw){\line(1,0){\xw}}
\put(\xw,  0){\line(0,1){\yw}}
\put(  0,  0){\line(0,1){\yw}} }
\end{picture} 
\end{figure}  

\clearpage

\begin{figure}[t] 
\unitlength 2.0mm
\begin{picture}(60,60)(-10,-10)
\def\xw{60.000000} \def\yw{40.000000}
\put(22,-20){\Huge Figure 3(a)}
\put(26,-8){\huge $T/T_c$}
\put(-10,43){\huge $M_{\eta'}(T)/M_{\eta'}(0)$}
\def\errorbar2#1#2#3#4#5#6{
\put(#1,#2){\line(0,1){#3}}
\put(#1,#2){\line(0,-1){#3}}
\put(#4,#5){\line(1,0){1}}
\put(#4,#6){\line(1,0){1}} }
\errorbar2{0.000000}{40.000000}{0.000000}{-0.500000}{40.000000}{40.000000}
\errorbar2{45.000000}{40.000000}{0.800000}{44.500000}{40.800000}{39.200000}
\errorbar2{55.799999}{38.000000}{0.400000}{55.299999}{38.400000}{37.600000}
\put(0.000000,40.000000){\circle*{1.000000}}
\put(45.000000,40.000000){\circle*{1.000000}}
\put(55.799999,38.000000){\circle*{1.000000}}
\def\errorbar2#1#2#3#4#5#6{
\put(#1,#2){\line(0,1){#3}}
\put(#1,#2){\line(0,-1){#3}}
\put(#4,#5){\line(1,0){1}}
\put(#4,#6){\line(1,0){1}} }
\errorbar2{0.000000}{40.000000}{0.000000}{-0.500000}{40.000000}{40.000000}
\errorbar2{45.000000}{34.400002}{0.800000}{44.500000}{35.200002}{33.600002}
\errorbar2{55.799999}{27.600000}{0.400000}{55.299999}{28.000000}{27.200000}
\put(0.000000,40.000000){\circle{1.000000}}
\put(45.000000,34.400002){\circle{1.000000}}
\put(55.799999,27.600000){\circle{1.000000}}
\put(0.000000,0){\line(0,1){1}}
\put(12.000000,0){\line(0,1){1}}
\put(24.000000,0){\line(0,1){1}}
\put(36.000000,0){\line(0,1){1}}
\put(48.000000,0){\line(0,1){1}}
\put(60.000000,0){\line(0,1){1}}
\put(0.000000,0){\line(0,1){1}}
\put(-2.200000,-3.5){\huge 0.0}
\put(9.800000,-3.5){\huge 0.2}
\put(21.800000,-3.5){\huge 0.4}
\put(33.800000,-3.5){\huge 0.6}
\put(45.800000,-3.5){\huge 0.8}
\put(57.800000,-3.5){\huge 1.0}
\put(1.100000,-3.5){\huge }
\put(0,0.000000){\line(1,0){1}}
\put(\xw,0.000000){\line(-1,0){1}}
\put(0,8.000000){\line(1,0){1}}
\put(\xw,8.000000){\line(-1,0){1}}
\put(0,16.000000){\line(1,0){1}}
\put(\xw,16.000000){\line(-1,0){1}}
\put(0,24.000000){\line(1,0){1}}
\put(\xw,24.000000){\line(-1,0){1}}
\put(0,32.000000){\line(1,0){1}}
\put(\xw,32.000000){\line(-1,0){1}}
\put(0,40.000000){\line(1,0){1}}
\put(\xw,40.000000){\line(-1,0){1}}
\put(0,0.000000){\line(1,0){1}}
\put(\xw,0.000000){\line(-1,0){1}}
\put(-5.700000,-1.000000){\huge 0.0}
\put(-5.700000,6.900000){\huge 0.2}
\put(-5.700000,14.900000){\huge 0.4}
\put(-5.700000,22.900000){\huge 0.6}
\put(-5.700000,30.900000){\huge 0.8}
\put(-5.700000,38.900000){\huge 1.0}
{\linethickness{0.25mm}
\put(  0,  0){\line(1,0){\xw}}
\put(  0,\yw){\line(1,0){\xw}}
\put(\xw,  0){\line(0,1){\yw}}
\put(  0,  0){\line(0,1){\yw}} }
\end{picture} 
\end{figure}  

\clearpage

\begin{figure}[t] 
\unitlength 2.0mm
\begin{picture}(60,60)(-10,-10)
\def\xw{60.000000} \def\yw{40.000000}
\put(22,-20){\Huge Figure 3(b)}
\put(26,-8){\huge $T/T_c$}
\put(-10,43){\huge $M_{\eta}(T)/M_{\eta}(0)$}
\def\errorbar2#1#2#3#4#5#6{
\put(#1,#2){\line(0,1){#3}}
\put(#1,#2){\line(0,-1){#3}}
\put(#4,#5){\line(1,0){1}}
\put(#4,#6){\line(1,0){1}} }
\errorbar2{0.000000}{40.000000}{0.000000}{-0.500000}{40.000000}{40.000000}
\errorbar2{45.000000}{40.000000}{0.800000}{44.500000}{40.800000}{39.200000}
\errorbar2{55.799999}{39.599998}{0.400000}{55.299999}{39.999998}{39.199998}
\put(0.000000,40.000000){\circle*{1.000000}}
\put(45.000000,40.000000){\circle*{1.000000}}
\put(55.799999,39.599998){\circle*{1.000000}}
\def\errorbar2#1#2#3#4#5#6{
\put(#1,#2){\line(0,1){#3}}
\put(#1,#2){\line(0,-1){#3}}
\put(#4,#5){\line(1,0){1}}
\put(#4,#6){\line(1,0){1}} }
\errorbar2{0.000000}{40.000000}{0.000000}{-0.500000}{40.000000}{40.000000}
\errorbar2{45.000000}{38.800003}{0.800000}{44.500000}{39.600003}{38.000003}
\errorbar2{55.799999}{34.000000}{0.400000}{55.299999}{34.400000}{33.600000}
\put(0.000000,40.000000){\circle{1.000000}}
\put(45.000000,38.800003){\circle{1.000000}}
\put(55.799999,34.000000){\circle{1.000000}}
\put(0.000000,0){\line(0,1){1}}
\put(12.000000,0){\line(0,1){1}}
\put(24.000000,0){\line(0,1){1}}
\put(36.000000,0){\line(0,1){1}}
\put(48.000000,0){\line(0,1){1}}
\put(60.000000,0){\line(0,1){1}}
\put(0.000000,0){\line(0,1){1}}
\put(-2.200000,-3.5){\huge 0.0}
\put(9.800000,-3.5){\huge 0.2}
\put(21.800000,-3.5){\huge 0.4}
\put(33.800000,-3.5){\huge 0.6}
\put(45.800000,-3.5){\huge 0.8}
\put(57.800000,-3.5){\huge 1.0}
\put(1.100000,-3.5){\huge }
\put(0,0.000000){\line(1,0){1}}
\put(\xw,0.000000){\line(-1,0){1}}
\put(0,8.000000){\line(1,0){1}}
\put(\xw,8.000000){\line(-1,0){1}}
\put(0,16.000000){\line(1,0){1}}
\put(\xw,16.000000){\line(-1,0){1}}
\put(0,24.000000){\line(1,0){1}}
\put(\xw,24.000000){\line(-1,0){1}}
\put(0,32.000000){\line(1,0){1}}
\put(\xw,32.000000){\line(-1,0){1}}
\put(0,40.000000){\line(1,0){1}}
\put(\xw,40.000000){\line(-1,0){1}}
\put(0,0.000000){\line(1,0){1}}
\put(\xw,0.000000){\line(-1,0){1}}
\put(-5.700000,-1.000000){\huge 0.0}
\put(-5.700000,6.900000){\huge 0.2}
\put(-5.700000,14.900000){\huge 0.4}
\put(-5.700000,22.900000){\huge 0.6}
\put(-5.700000,30.900000){\huge 0.8}
\put(-5.700000,38.900000){\huge 1.0}
{\linethickness{0.25mm}
\put(  0,  0){\line(1,0){\xw}}
\put(  0,\yw){\line(1,0){\xw}}
\put(\xw,  0){\line(0,1){\yw}}
\put(  0,  0){\line(0,1){\yw}} }
\end{picture} 
\end{figure}  

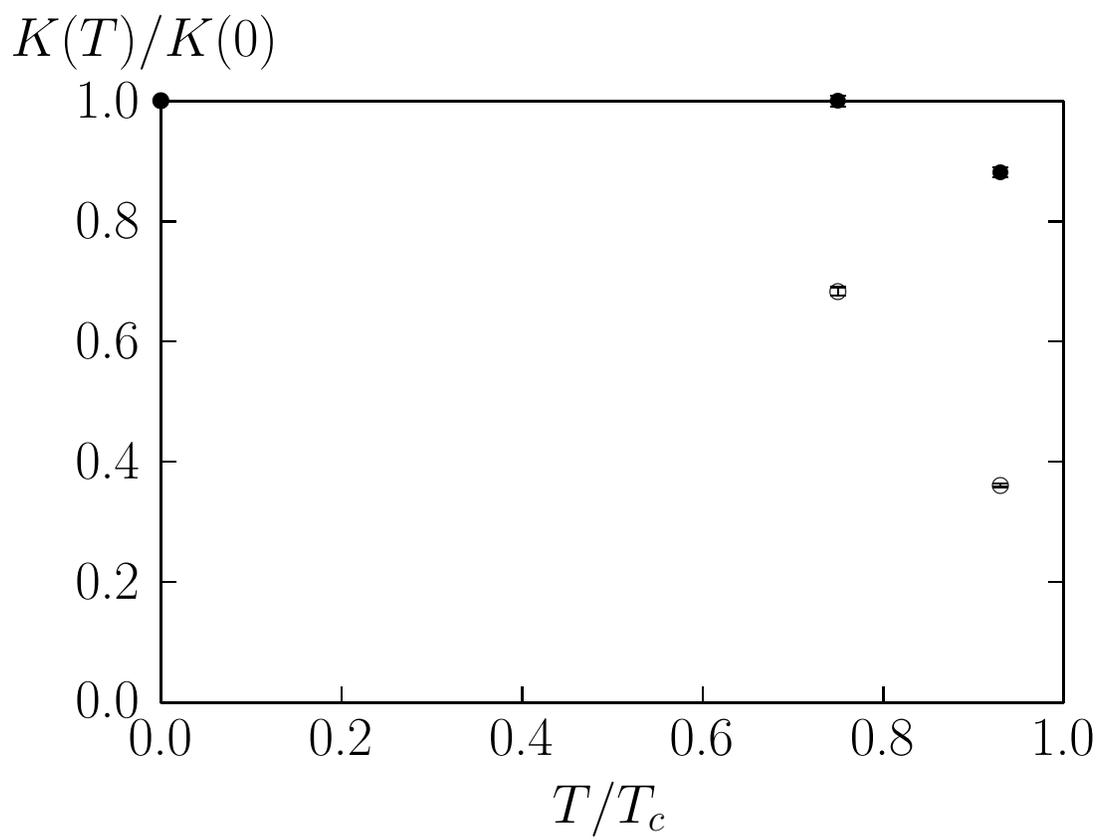
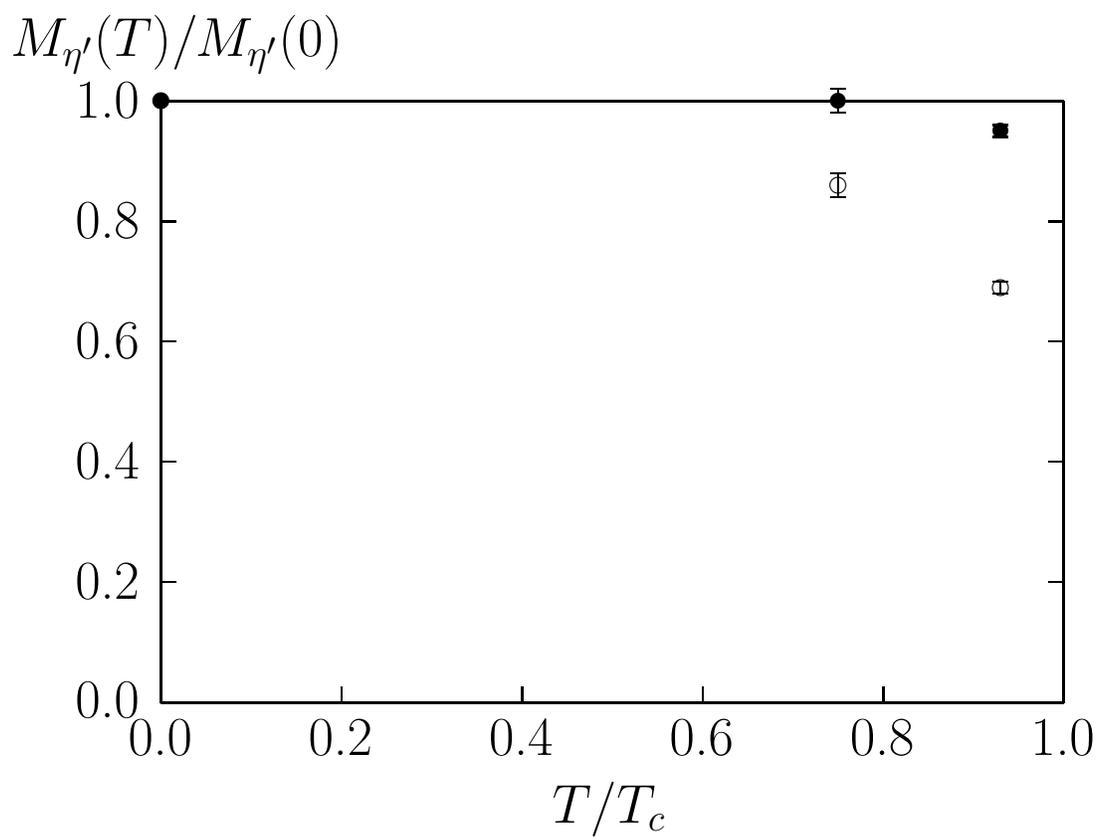
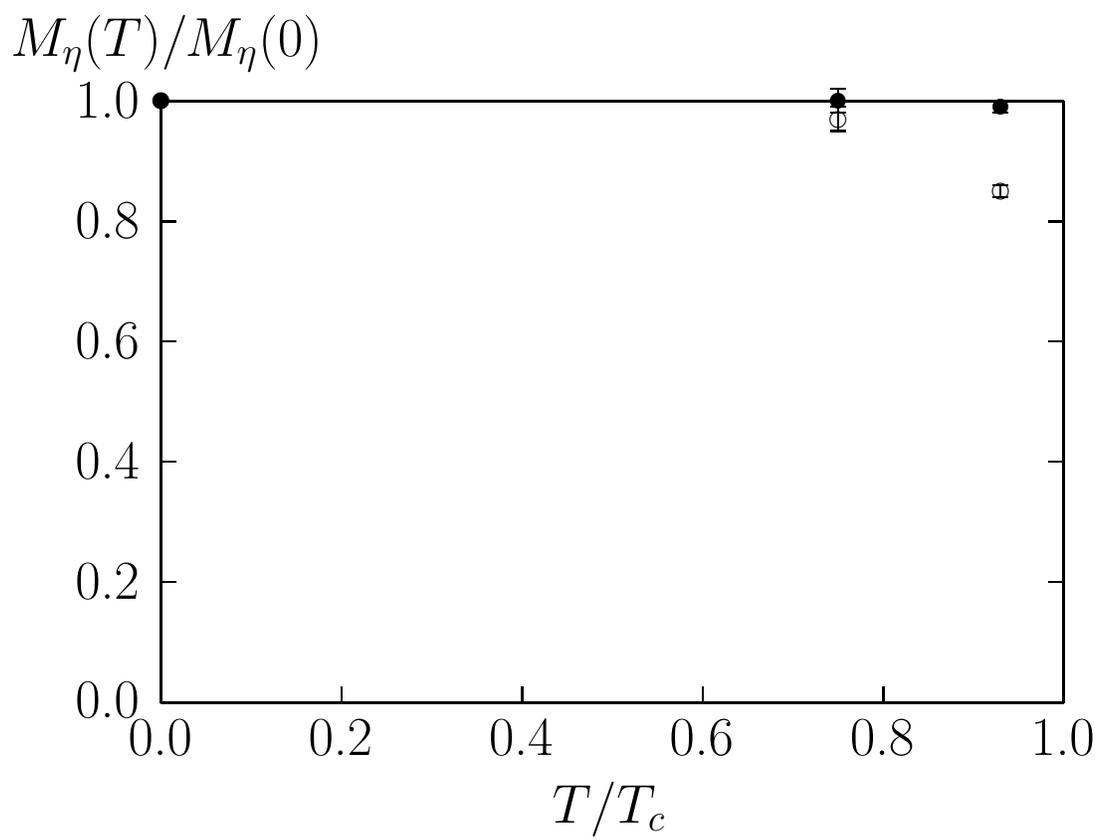
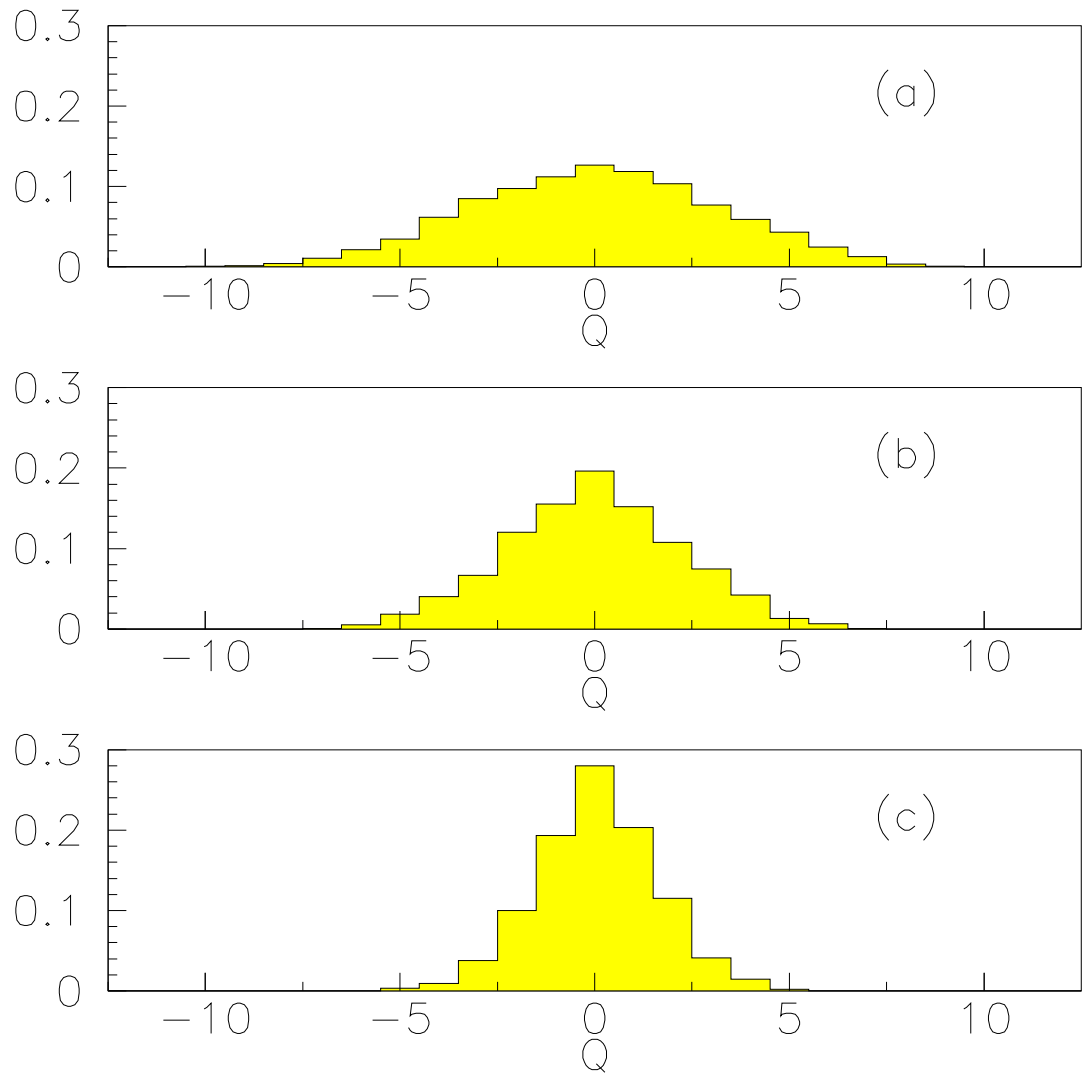
\end{document}